\documentclass[prl,floatfix,twocolumn,preprintnumbers,amsmath,amssymb,superscriptaddress]{revtex4}
\usepackage{graphicx,psfrag}
\usepackage{amsfonts}

\newcommand{\newc}{\newcommand}
\newc{\gsim}{\lower.7ex\hbox{$\;\stackrel{\textstyle>}{\sim}\;$}}
\newc{\lsim}{\lower.7ex\hbox{$\;\stackrel{\textstyle<}{\sim}\;$}}
\newc{\gev}{\,{\rm GeV}}
\newc{\mev}{\,{\rm MeV}}
\newc{\ev}{\,{\rm eV}}
\newc{\kev}{\,{\rm keV}}
\newc{\tev}{\,{\rm TeV}}

\newc{\mz}{M_Z}
\newc{\mpl}{M_*}
\newc{\mw}{m_{\rm weak}}
\newc{\nr}[1]{N^c_R{}_{#1}}
\def\beq{\begin{equation}}
\def\eeq{\end{equation}}
\def\bea{\begin{eqnarray}}
\def\eea{\end{eqnarray}}
\def\bitem{\begin{itemize}}
\def\eitem{\end{itemize}}
\newc{\ie}{{\it i.e.}}          \newc{\etal}{{\it et al.}}
\newc{\eg}{{\it e.g.}}          \newc{\etc}{{\it etc.}}
\newc{\cf}{{\it c.f.}}
\def\bar#1{\overline{#1}}

\def\abs#1{\left| #1\right|}

\def\inv{^{\raise.15ex\hbox{${\scriptscriptstyle -}$}\kern-.05em 1}}
\def\lbar{{\lower.35ex\hbox{$\mathchar'26$}\mkern-10mu\lambda}} %lambda bar

\let\<=\langle
\let\>=\rangle

\let\+=\uparrow

\let\ga=\gamma

\let\De=\Delta

\let\La=\Lambda

\let\si=\sigma

\let\Om=\Omega

\begin{document}    
\thispagestyle{empty}

\begin{center}
\title{Light WIMPs in the Sun: Constraints from Helioseismology}
\date[]{Revised version resubmitted to Phys.\ Rev.\  D: 3rd September 2010. Accepted for publication:  23rd September, 2010}
\author{Daniel T. Cumberbatch}
\email[]{D.Cumberbatch@sheffield.ac.uk}
\affiliation{Astroparticle Theory \& Cosmology Group, Department of Physics \& Astronomy, The University of Sheffield, Hicks Building, Hounsfield Road, Sheffield, S3 7RH, U.K}
\author{Joyce. A.  Guzik}
\email[]{joy@lanl.gov}
\affiliation{Los Alamos National Laboratory, XTD-2 MS T086, Los Alamos, NM 87545-2345 USA}
\author{Joseph Silk}
\email[]{j.silk1@physics.ox.ac.uk}
\affiliation{Physics Department, University of Oxford, Oxford, OX1 3RH, UK}
\author{L. Scott Watson}
\email[]{LSWATSO@SANDIA.GOV}
\affiliation{Sandia National Laboratories, PO Box 5800 MS-0431, Albuquerque, NM 87185, USA }
\author{Stephen M. West}
\email[]{stephen.west@rhul.ac.uk}
\affiliation{Royal Holloway, University of London, Egham, TW20 0EX, UK}
\affiliation{Rutherford Appleton Laboratory, Chilton, Didcot, OX11 0QX, UK}

%***************************************************************************************************
\begin{abstract}
\noindent  We calculate solar models including dark matter (DM) weakly-interacting massive particles (WIMPs) of mass 5-50\gev\;and test these models against helioseismic constraints on sound speed, convection zone depth, convection zone helium abundance, and small separations of low-degree p-modes.   Our main conclusion is that both direct detection experiments and particle accelerators may be complemented by using the Sun as a probe for WIMP DM particles in the 5-50\gev\;mass range. The DM most sensitive to this probe has suppressed annihilations and a large spin-dependent elastic scattering cross section.
For the WIMP cross-section parameters explored here, the lightest WIMP masses $<$\,10\gev\;are ruled out by constraints on core sound speed and low-degree frequency spacings.  For WIMP masses 30-50\gev, the changes to the solar structure are confined to the inner 4\% of the solar radius and so do not significantly affect the solar p-modes. Future helioseismology observations, most notably involving g-modes, and future solar neutrino experiments may be able to constrain the allowable DM parameter space in a mass range that is of current interest for direct detection. 
\end{abstract}

\maketitle 
\end{center}

%***************************************************************************************************

\section{Introduction}
\noindent
We explore the role of DM WIMPs in modifying the thermal gradient of the Sun. A similar study involving standard WIMPS of mass $50\gev$  or larger has been performed by Bottino et al. \cite{Bottino}. However, here we wish to consider the effects of low mass WIMPs, with masses as low as $5\gev$, with large trapped abundances within the Sun. 

To affect the Sun's thermal gradient, we need large elastic scattering rates. The solar sound speed can be  affected, and helioseismology has been  proposed as providing a possible constraint on supersymmetric WIMPs \cite{lopesa,lopesb, lopesc}. For the range in masses in which we are interested, the limits on the size of spin-independent WIMP elastic cross sections from CRESST \cite{cresst}, XENON-10 \cite{xenon10}, and XENON-100 \cite{xenon100} are already quite stringent making it unlikely for helioseismology to provide further restrictions. 
Moreover, recent COUPP results \cite{Behnke} (especially at masses $\gsim 10$\gev) and the results from PICASSO \cite{picasso} similarly restrict spin-dependent interactions. However, these limits become weaker as the DM mass is decreased, especially for masses of around $5$\gev or less.

Detailed solar models \cite{Bottino} show that a WIMP signal is only possible for cross sections in a limited range. 
For a DM particle mass of $50 \gev$ the relevant effective cross section for these signals was found to be of order $10^{-35}\,{\rm cm}^2$. The limits on spin-dependent scattering from direct detection experiments, e.g., COUPP \cite{Behnke, coupp} restrict the spin-dependent elastic scattering cross section for a $50 \gev$ DM particle to less than $\sim10^{-37}\,{\rm cm}^2$.  Moving to lighter masses alleviates these limits somewhat, and for around $5 \gev$ we are in the interesting region of around $10^{-35}\,{\rm cm}^2$ \cite{picasso}. In addition, given the astrophysical uncertainties that can affect these limits (e.g.\,\cite{green}) it is intriguing to ask whether helioseismology can complement direct detection limits at lower masses.

Recently, there has been interest in exploring the low DM mass regime as a possible way to consistently combine the results from the DAMA direct detection experiment \cite{dama} with those from others such as CDMS \cite{cdms} and CoGeNT \cite{cogent} (see e.g., \cite{lowmasssector}). Although we do not attempt to do the same here, we simply note that this mass regime is of great interest with upper limits on the spin-dependent elastic scattering cross section for low-mass DM, reaching $\sim10^{-32}\,{\rm cm}^2$ \cite{Kopp} for certain models and assumptions.

Solar effects are most pronounced for DM with a suppressed annihilation cross section such that after capture by the Sun, the DM candidates do not annihilate quickly. A prominent example of this is asymmetric DM where annihilation is completely suppressed. 

In this paper, we outline a class of models for WIMPs that are capable of modifying the temperature profile in the core of the Sun, and illustrate their effects on helioseismology and neutrino fluxes. The accumulation of these WIMPs in the solar core results in significant energy transfer to solar protons. We note that the effect is not large enough to account for discrepancies between observations and helioseismology for models that also predict the observed neutrino flux.  However, given the current debate about the appropriate element abundances to be adopted in solar models \cite{serenelli}, this effect may still play a role and should be included in the models.  Indeed the recently revised solar abundances result in solar models that cannot reproduce currently observed helioseismic data \cite{AGS05, AGSS09}.

The effects on the Sun of low-mass asymmetric WIMPs possessing large self-interactions have been considered in \cite{Frandsen}. While in \cite{Frandsen} the authors focus on WIMPs with spin-independent interactions, in this study we specifically focus on WIMPs with purely axial interactions and consequently only spin-dependent elastic scattering.

In the following sections, we outline the main features of DM models that can be potentially probed using solar properties and explore their effect on solar models. 

%***************************************************************************************************

\section{WIMP Model Characteristics}

The most interesting DM mass region is the low mass region around $10\gev$ and below. As stated in the introduction, direct detection constraints restrict the size of the DM elastic scattering cross section. In particular, the constraints on spin-independent elastic scattering for low masses by CRESST \cite{cresst} are too stringent for any possible improvements from helioseismology. We will therefore focus on WIMPs that dominantly undergo spin-dependent scattering, the limits on which are less stringent.

We consider any model that can give rise to the following effective axial interaction
%%%%%%%%%%%%%%%%%%%%%%%%%%%%%%%%%%%%%%%%%%%%%%%
\beq
\mathcal{L}\supset \sum_q\frac{1}{\La_q^2}\bar{\chi}\ga^{\mu}\ga_5\chi\bar{q}\ga_{\mu}\ga_5q,
\label{eq:effint}
\eeq
%%%%%%%%%%%%%%%%%%%%%%%%%%%%%%%%%%%%%%%%%%%%%%%
between our DM particle, $\chi$, and standard model quarks, $q$, where $\La_q$ is a dimensionful parameter encoding the effective energy scale of the dynamics that generate this interaction. 

The $\chi-p$ spin-dependent elastic scattering cross section at zero momentum transfer, resulting from the purely axial vector interaction, is given in \cite{Gondolo:2004sc} as
%%%%%%%%%%%%%%%%%%%%%%%%%%%%%%%%%%%%%%%%%%%%%%
\beq
\si^{\rm\bf sd}_{\chi p}=\frac{3m_\chi^2m_p^2}{\pi (m_\chi+m_p)^2}\abs{\frac{1}{\La^2_u} \De u+\frac{1}{\La^2_d} \De d+\frac{1}{\La^2_s} \De s}^2,
\eeq
%%%%%%%%%%%%%%%%%%%%%%%%%%%%%%%%%%%%%%%%%%%%%%
where the $\De q$ factors are the spin fractions of the proton carried by a given quark \cite{Gondolo:2004sc, bhs, Bass:2004xa}. It is this spin-dependent cross section that is central to our analysis.

In addition to spin-dependent scattering, the DM model must have a suppressed annihilation cross section, compared to that required by successful DM genesis via thermal freeze-out, in order for DM particles to continuously accumulate in the Sun. The minimum suppression required for an effect to be seen is of the order of a p-wave suppression compared to the s-wave annihilation rate required by standard freeze-out. Therefore, a suppression of order $b/a$, where $\langle\sigma_{\rm ann.}\upsilon\rangle=a+b\upsilon^2+O(\upsilon^4)$, is required in the standard small velocity expansion of the WIMP annihilation cross section, where $\upsilon$ is the relative velocity between two colliding WIMP DM particles and the angled brackets represent a thermal average.

A suppressed annihilation rate is not a generic feature of models of thermal freeze-out. However, models where the DM species possesses a particle-antiparticle asymmetry can lead to DM with zero annihilations today. The relic abundance in this case is assumed to be fixed by the value of the asymmetry, with the ratio of the baryon to DM relic abundance, $\Om_{\rm b}/\Om_{\chi}$, determined by the dynamics that generate the asymmetry and is of order $m_p/m_{\chi}$. A number of attempts to link $\Om_{\rm b}$ to $\Om_{\chi}$ have been made in the context of asymmetric DM (see e.g., \cite{asymm}). Typically, models of asymmetric DM involve DM particles with small masses and could therefore have implications for helioseismology. Previously, models of DM possessing large spin-dependent elastic scattering cross-section and an asymmetry have been investigated in the context of their effects on neutrino fluxes produced in the Sun \cite{gr}. Further related ideas can be found in \cite{solarneuprob}.

%***************************************************************************************************

\section{WIMP capture in the Sun}

The accretion rate of the number of WIMPs captured by the Sun in the 
large spin-dependent scattering cross section limit, where all WIMPs intercepting the Sun are captured, is given by  \cite{gould1987}
%%%%%%%%%%%%%%%%%%%%%%%%%%%%%%%%%%%%%%%%%%%
\begin{equation}
\Gamma= \left(\frac{8}{3\pi}\right)^{1/2}
\frac{\rho_{\rm DM}}{m_\chi} \bar v
\left[\zeta + \frac{3 v_{\rm esc.}^2}{2\bar v^2}\right]
\xi(\infty)\pi R_\odot^2,
\end{equation}
%%%%%%%%%%%%%%%%%%%%%%%%%%%%%%%%%%%%%%%%%%%
where $\upsilon_{\rm esc.}\approx 617\,{\rm km\,s}^{-1}$ is the escape velocity at the solar surface, $\bar \upsilon \approx 270\,{\rm km\,s}^{-1}$, $\xi(\infty)\approx0.75,$ and $\zeta =1.77.$
This reduces to
%%%%%%%%%%%%%%%%%%%%%%%%%%%%%%%%%%%%%%%%%%%
\begin{equation}
\frac{3.042 \times 10^{25}}{ m_{\chi} (\rm GeV)} \frac{\rho_{\rm DM}}{0.3\,{\rm GeV\,cm}^{-3}}\,\,s^{-1},
\end{equation}
%%%%%%%%%%%%%%%%%%%%%%%%%%%%%%%%%%%%%%%%%%%%
assuming that the WIMP interaction cross section with matter is $\sigma_{\chi}\gtrsim10^{-36}\,{\rm cm}^2$ in order for all incoming WIMPs to undergo one or more scatterings while inside the Sun. We normalise the local DM density in the solar neighbourhood to $\rho_{\rm DM}=0.3\,{\rm GeV\,cm}^{-3}$.  Here, $m_\chi$ is the WIMP mass. Because of competing effects of annihilation and evaporation, the number of accreted WIMPs at time $t$ is obtained by solving the differential equation
%%%%%%%%%%%%%%%%%%%%%%%%%%%%%%
\begin{equation}
{\dot N}= {\cal F}-\Gamma_{\rm ann.}-\Gamma_{\rm evap.}, 
\end{equation}
%%%%%%%%%%%%%%%%%%%%%%%%%%%%%%
where $\Gamma_{\rm ann.}$ is the self-annihilation rate $\Gamma_{\rm ann.}=\langle\sigma_{\rm ann.} \upsilon\rangle \int n^2_{\chi} {\rm d}V=\frac{\langle\sigma_{\rm ann.} \upsilon\rangle n^2_{\chi}V^{\rm 2}}{V}=C_{\rm A}N(t)^{\rm 2}$.  Here, $\langle\sigma_{\rm ann.}\upsilon\rangle$ is the product of thermally-averaged WIMP self-annihilation cross section and velocity, and  $n_\chi=\frac{\rho_{\rm DM}}{m_{\chi}}$, the number density of WIMPs inside the Sun, is assumed to be constant. The evaporation rate, $\Gamma_{\rm evap.}$, decays exponentially with temperature as $\sim e^{-G M m_{\chi}/R T}$  and is negligible with respect to the annihilation rate for  
$m_\chi \gtrsim10\gev$. With this simplification, the population of WIMPs at time $t$ is given by
%%%%%%%%%%%%%%%%%%%%%%%%%
\begin{equation}
N(t)= ({\cal F} \tau)\; {\rm tanh}(t/ \tau)/2,
\end{equation}
%%%%%%%%%%%%%%%%%%%%%%%%%%
where the time-scale $\tau=1/{\rm \sqrt{{\cal F }C_{\rm A}}}$. For $t\gg\tau$, i.e. when the equilibrium between accretion and annihilation has been reached, the number of particles accreted is equal to the time-independent product ${\mathcal F}\tau$.
  
Assuming that we are in the regime when the velocities and positions of the WIMPs within the Sun follow a Maxwell-Boltzmann distribution with respect to its centre, the amount of energy released by annihilation in a thermalisation volume centred in the compact star will have a radius \cite{griest1987}
%%%%%%%%%%%%%%%%%%%%%%%%%%%%%%%%%%%%%%%%%%%%%%
\begin{eqnarray}
R_{\rm th.}&=&\left( \frac{9 k_{\rm B}T_c}{4 \pi G \rho_c m_{\chi}}\right)^{1/2}\nonumber\\
&=&6.4\times10^9\,{\rm cm}\left(\frac{T}{10^{\rm 7}\,{\rm K}}\right)^{1/2}\left(\frac{1\,{\rm g\,cm^{-3}}}{\rho_{\rm c}}\right)^{1/2}\nonumber\\
&&\times\left(\frac{100\,{\rm GeV}}{m_{\chi}}\right)^{1/2}.
\end{eqnarray}
%%%%%%%%%%%%%%%%%%%%%%%%%%%%%%%%%%%%%%%%%%%%%%
Taking typical solar core  conditions as having central internal temperature $T_c=1.57\times 10^{\rm 7}$\,K and density $\rho_{\rm c}$=154\,{\rm g\,cm$^{-3}$, we find that 
$R_{\rm th.}\simeq0.03\left(m_\chi/10\gev\right )^{-1/2}\,R_\odot$.  We assume constant temperature and density within the WIMP thermalisation volume. Hence, WIMPs fill the solar core for masses in the range $\sim$5-10\gev, and it is for this reason that there may be a significant imprint on helioseismology for both p- and g-modes that are sensitive to the sound crossing time across the solar interior. At lower masses, evaporation predominates, while at higher masses, the WIMPs within the Sun occupy a smaller volume.

%***************************************************************************************************

\section{Application to solar models}
\noindent The solar models shown here are evolved from the pre-main sequence using an updated version of the one-dimensional evolution codes described in 
\citet{Iben_1963, Iben_1965a, Iben_1965b}. The evolution code uses the SIREFF EOS \citep[see][]{GS97}, Burgers (1969) \cite{burgers69} diffusion treatment as implemented by \cite{CGK89, IM85}, the nuclear reaction rates from \citet{Angulo_1999} with a correction to the $^{14}$N rate from \citet{Formicola_2004}, and the OPAL opacities \citep{IR96} supplemented by the \citet{Fer05} low-temperature opacities.

The models are calibrated to the present solar radius $R_{\odot}=6.9599\times10^{10}$\,cm \citep{Allen_1973}, luminosity $L_{\odot}=3.846\times10^{33}$\,erg\,s$^{-1}$ \citep{Willson_1986}, mass $M_{\odot}=1.989\times10^{33}$\,g \citep{CT86}, and age 4.54$\,\pm\,$0.04\,Gyr \citep{Guenther_1992}. Defining $X$ and $Z$ as the mass fraction of hydrogen, and the mass fraction of elements heavier than helium, respectively, the models are calibrated to the photospheric $Z/X$ ratio appropriate for either the Asplund, Grevesse and Sauval (2005) solar mixture \cite{AGS05} (hereafter AGS05), or the Grevesse and Noels (1993) solar mixture \cite{GN93} (hereafter GN93). 

For the evolution models, the initial helium abundance, $Y_0$, initial heavy element mass fraction, $Z_0$, and mixing length to pressure-scale-height ratio, $\alpha$ are adjusted so that the final luminosity, radius, and surface $Z/X$ match the above constraints to within uncertainties.

From the final evolution model, a more finely zoned model is created for calculating the oscillation frequencies. The radial and non-radial non-adiabatic p-mode and g-mode frequencies are calculated using the Lagrangian pulsation code developed by Pesnell \cite{Pesnell_1990}. (See \cite{GWC05} for additional references and description of the physics used in the evolution and pulsation codes and models.)

The WIMP energy transport description is considered in two regimes, depending critically on the mean free path of the WIMPs and the scale radius of the system. This ratio is known as the Knudsen parameter,
%%%%%%%%%%%%%%%%%%%%%%%%%
\begin{equation}
K_{\rm n}=\frac{l_{\chi,i}(r)}{r_{\chi}},
\end{equation}
%%%%%%%%%%%%%%%%%%%%%%%%%
where $l_{\chi,i}(r)$, the mean free path of the WIMPs relative to the $\it i$th element, and $r_{\chi}$, the WIMP scale radius are defined, respectively, as 
%%%%%%%%%%%%%%%%%%%%%%%%%
\begin{equation}
\label{eqn:MFP}
l_{\chi,i}(r)=\sum_i\;\frac{m_i}{\sigma_iX_i(r)\rho(r)},
\end{equation}
%%%%%%%%%%%%%%%%%%%%%%%%%
where $\si_{i}$ is the elastic scattering cross section, $X_i(r)$ is the mass fraction of isotopic species $i$ at radius $r$ (e.g. $X_1$ is the proton), $\rho(r)$ is the matter density in units of g\,cm$^{-3}$ at radius $r$, and $m_i$ is the mass of species $i$ (e.g. the proton mass for hydrogen)
and
%%%%%%%%%%%%%%%%%%%%%%%%%
\begin{equation}
r_{\chi} \,= \, \left(\frac{3 k_{\rm B} T_{\chi}}{2\pi G \rho_c m_{\chi}}\right)^{1/2},
\end{equation}
%%%%%%%%%%%%%%%%%%%%%%%%
where $\rho_c$ is the central solar density, $T_{\chi}$ is the WIMP temperature, $k_{\rm  B}$ is Boltzmann's constant, $G$ is the gravitational constant, and $m_{\chi}$ is the WIMP mass.

For the models considered here, only spin-dependent interactions are considered, with the contribution of hydrogen overwhelming the spin interactions of the core. Thus, all formulas that are summed over the elements can be reduced to their hydrogen nuclei (i.e. proton) contribution.

In \citep{GR90a, GR90b}, a Monte Carlo method was used to solve several ``generic toy star" models incorporating WIMPs in the conductive regime. The authors found that by using the conduction formula along with a suppression factor related to the Knudsen parameter, the entire non-local transport regime could be related to the conductive regime through the approximation
%%%%%%%%%%%%%%%%%%%%%%%%
\begin{equation}
\label{eqn:GR90}
L_x(r) ~=~ f(K_{\rm n})L_{\rm cond.}(r)
\end{equation}
%%%%%%%%%%%%%%%%%%%%%%%%
where $L_x$ is the total energy transferred from WIMPs to the nuclei in the intermediate regime between the conductive and non-local regimes, $L_{\rm cond.}$ is the energy transported by WIMPs in the conductive regime and the suppression factor $f(K_{\rm n})$ is given by
%%%%%%%%%%%%%%%%%%%%%%%%%%%%%%%%
\begin{equation}
f(K_{\rm n}) ~=~ \frac{1}{\left(\frac{K_{\rm n}}{K_o}\right)^2 \, +1}
\end{equation}
%%%%%%%%%%%%%%%%%%%%%%%%%%%%%%%%%
The Knudsen number $K_o$ is the mean free path, in scale height units, that gives the most efficient energy transport from the WIMPs within the Sun to the surrounding nuclei. Gould \& Raffelt found this to be equal to $\sim$\,0.4 \citep{GR90a, GR90b}.
For the solar models considered here, WIMPs are introduced
into the energy transport by modifying the opacity 
%%%%%%%%%%%%%%%%%%%%%%%%%%%%%%%%%%
\begin{equation}
\label{eqn:wimpopacity}
\frac{1}{\kappa_{\rm total}}\,=\,\frac{1}{\kappa_{{\rm rad.}+e^-}}\,+\,\frac{f(K_{\rm n})}{\kappa_{\rm  cond.}}.
\end{equation}
%%%%%%%%%%%%%%%%%%%%%%%%%%%%%%%%%%%

\noindent
Here $\kappa_{{\rm rad.}+e^-}$ is the combination of Rosseland mean radiative opacity and an effective opacity to take into account electron thermal conduction that are added in reciprocal; ${\kappa_{\rm  cond.}}$ is the effective opacity derived by treating WIMP energy transport as a conductive process.

In the introduction, we noted that for DM masses heavier than $5-10\gev$, direct detection places constraints on the spin-dependent elastic scattering cross section. However, here we want to demonstrate and compare the effects of our described DM particles on the Sun as we decrease their mass to values where their corresponding spin-dependent elastic scattering cross sections are permitted by direct detection limits. 

We have also explored models with WIMPs of different masses and interaction cross sections \cite{WatsonDissertation}, and present this series as an illustration of the effects of WIMPs on a solar model when one parameter, namely, the WIMP mass, is varied.  
We chose an interaction cross section, $\sigma_x$= 7$\times10^{-35}$ cm$^2$, and a very small annihilation cross section, $\langle\sigma_{\rm ann.}\upsilon\rangle$\,=\,10$^{-40} \,{\rm cm}^3\,{\rm s}^{-1}$, to enhance the effect of WIMPS on the solar model.  
These results are intended to explore whether helioseismic signatures could have the potential to reveal or rule out the presence of a particular class of WIMP. At this stage we do not intend to provide rigorous helioseismic constraints on the properties of WIMPs.  To this end, we compare the characteristic parameters of solar models based on the GN93 and AGS05 solar abundances when including WIMPs of masses 50, 30, 20, 15, 10, and 5\,$\gev$. 

The solar models including WIMPs use a tiny but non-zero annihilation rate. This is in contrast to the DM models outlined above where, due to the asymmetry in the DM species, the DM particles are unable to annihilate once captured by the Sun. This means that the annihilation rate used in our numerical solar models should also be zero. It turns out that the size of the effects manifesting in solar properties plateaus such that the effect of decreasing the annihilation rate further does not significantly change the numerical results \cite{Bottino}. 

%***************************************************************************************************

\subsection{Effect on model structure}
%%%%%%%%%%%%%%%%%%%%%%%%%%%%%%%%%%%%%%%%%%%%%%%
\begin{table*}[!t] 
\begin{minipage}{\textwidth}
\caption[Caption]%
{Properties of Standard Solar Models and Solar Models including WIMPs\footnote[1]{
Characteristic parameters of solar models based on the GN93 and AGS05 abundances, and AGS05 models including WIMPs of masses 50, 30, 20, 15, 10, and 5 GeV.  All models were run with $\left<\sigma_{\rm ann}v\right>$ = 1 x 10$^{-40}$ cm$^3$ s$^{-1}$ and a spin-dependent interaction cross section of $\sigma_x$ = 7 x 10$^{-35}$ cm$^2$ in order to enhanced the effects WIMPs have on the solar model.
}
}
\small
\begin{center}
\begin{tabular}{lcccccccc}
\noalign{\smallskip} \colrule\colrule \noalign{\smallskip}
Model/WIMP Mass:& GN93 & AGS05 &  50 GeV &  30 GeV  &  20 GeV  & 15 GeV & 10 GeV & 5 GeV\\
\noalign{\smallskip} \colrule
{\bf Model Calibration:}\footnote[2]{$X_0$, $Y_0$, and $Z_0$ are the initial mass fractions of hydrogen, helium and elements heavier than H and He, respectively; $\alpha$ is the mixing length to pressure-scale-height ratio; ZAMS is the zero-age main sequence.}\\
{}$X_0$& 0.71000 & 0.72950 & 0.72950 & 0.72950 & 0.72950 & 0.72950 & 0.72950 & 0.73570\\
{}$Y_0$& 0.27027 & 0.25698 & 0.25698 & 0.25698 & 0.25693 & 0.25690 & 0.25677 & 0.25057\\
{}$Z_0$& 0.01973 & 0.01352 & 0.01352 & 0.01352 & 0.01357 & 0.01360 & 0.01373 & 0.01373\\
{}$\alpha$& 2.0423 & 1.9916 & 1.9913 & 1.9910 & 1.9963 & 1.9990 & 2.0121 & 2.0734\\

Age-ZAMS (10$^9$ yrs)& 4.52 & 4.52 & 4.51 & 4.50 & 4.51 & 4.51 & 4.51& 4.51\\
log ($L/L_{\odot}$)& -5.74E-06 &  2.52E-06 &  -9.31E-06 & 7.86E-06 & -4.291E-06 & -7.16E-06 & -5.44E-06 & 2.5E-05\\
log ($R/R_{\odot}$)& 4.34E-07 & 2.17E-06  & 3.04E-06 & -1.74E-06 & 3.04E-06 & 8.69E-07 & 2.61E-06 & 3.04E-06\\
Z/X (surface)& 0.0246 & 0.01628 & 0.01628 & 0.01629 & 0.01635 & 0.01639 & 0.01657 & 0.01654\\
\\
{\bf Solar Center Properties at Solar Age:}\footnote[3]{$T_{\rm c}$, $\rho$$_{\rm c}$, $Y_{\rm c}$, $\kappa$$_{\rm c}$, and Sound Speed$_{\rm c}$ are the central temperature, density, helium mass fraction, opacity, and sound speed, respectively; the neutrino fluxes are given for the $^8$B and total fluxes at Earth's surface, in cm$^{-2} s^{-1}$ and in Solar Neutrino Units (SNUs) of 10$^{-36}$ absorptions per $^{37}$Cl atom per second.}\\
$T_{\rm c}$ (10$^{6}$\,K) & 15.64 & 15.42 & 15.172 & 15.005 & 14.814 & 14.646 & 14.330& 13.482\\
$\rho_{\rm c}$ (g\,cm$^{-3}$)& 152.40 & 148.96 & 149.84 & 150.70 & 152.45 & 154.11 & 158.31 & 175.57\\
$Y_{\rm c}$ & 0.6329 & 0.6183 & 0.6076 &0.6009& 0.5930 & 0.5851 & 0.5780 & 0.5488\\
$\kappa_{\rm c}$ (cm$^2$\,g$^{-1}$)& 1.231 & 1.261 & 0.02408 & 0.02254 & 0.02296 & 0.02290 & 0.02183 & 0.01756\\
Sound Speed$_{\rm c}$ (10$^7$\,cm\,s$^{-1}$)& 5.083 & 5.058 & 5.071& 5.061 & 5.050 & 5.040 & 5.007 & 4.94\\
$^8$B $\nu$ flux  (10$^6\,$cm$^{-2}$\,s$^{-1}$)& 5.26 & 4.30 & 4.19 & 3.98 & 3.68 & 3.30 & 2.60 & 1.04\\
$^8$B $\nu$ flux $^{37}$Cl detector (SNUs)& 5.99  & 4.91 & 4.77 & 4.53 & 4.19 & 3.76 & 2.96 & 1.18\\
Total $\nu$ flux $^{37}$Cl detector (SNUs)& 7.60 & 6.32 & 6.18 & 5.92 & 5.56 & 5.10 & 4.23 & 2.21\\
\\
{\bf Helioseismology:}\footnote[4]{The seismically-inferred CZ helium mass fraction and CZ base radius are 0.248 $\pm$ 0.003 and 0.713 $\pm$ 0.001 $R_{\odot}$, respectively \citep{BA04a}.}& & & &\\
$R_{\rm CZB}$ ($R_{\odot}$)& 0.7133 & 0.7294 & 0.7294 & 0.7293 & 0.7280 & 0.7280 & 0.7275& 0.7220\\
$Y_{\rm CZ}$& 0.2419 & 0.2273 & 0.2273 & 0.2273 & 0.2273 & 0.2273 & 0.2274 & 0.2227 \\
\noalign{\smallskip} \colrule
\end{tabular}
\end{center}
\normalsize
\label{tab:maxevo}
\end{minipage}
\end{table*}
%%%%%%%%%%%%%%%%%%%%%%%%%%%%%%%%%%%%%%%%%%%%%%%
\noindent In Table\,\ref{tab:maxevo}, we display the properties of standard solar models using either the GN93 or AGS05 abundances, as well as the properties of solar models when including WIMPs. 
Under the heading of ``Model Calibration'' we list the values of the following parameters: $X_0$, $Y_0$ and $Z_0$ are the initial mass fractions of H, He and metals (i.e. elements heavier than He); $\alpha$ is the mixing length-to-pressure scale height ratio;  ZAMS is the zero-age main sequence; {\texttt log}\,($L/L_{\odot}$) is the log luminosity in solar units; {\texttt log}\,($R/R_{\odot}$) is the log radius in solar units; Z/X (surface) is the surface ratio of metals to H mass fraction at the present solar age.  
Under ``Solar Center Properties'' we list the values of the following parameters: $T_{\rm c}$ is the central temperature; $\rho_{\rm c}$ is the central density; $\kappa_{\rm c}$ is the central opacity; $^8$B $\nu$ flux is the predicted $^8$B neutrino flux at Earth's surface, while the subsequent rows are the predicted total and $^8$B $\nu$ fluxes for $^{37}$Cl detectors, in Solar Neutrino Units (SNUs), defined as 10$^{-36}$ absorptions per $^{37}$Cl atom per second.   
Under the heading ``Helioseismology'', $R_{\rm CZB}$ is the predicted ratio of the convection zone base to the solar radius, $R_{\odot}$, and $ Y_{\rm CZ}$ is the predicted helium abundance in the convection zone.  The constraints on these quantities from helioseismology are given in the table end-notes.

Note that the model structure and calibration is considerably different for standard models without WIMPs calibrated to either the GN93 or AGS05 abundances.  Using the AGS05 abundances, which possess a smaller Z, a smaller helium mass fraction Y is required to compensate to increase the pressure in the core.  Since more hydrogen fuel is available, both $T_c$ and $\rho_c$ are slightly reduced to produce the same luminosity.  
The location of the envelope convection base is determined by the radius where the temperature gradient exceeds the adiabatic gradient.  For the AGS05 model, this point is reached at a lower temperature and larger radius because of the smaller fraction of heavier elements, particularly oxygen and neon, that are ionizing near the convection zone base and contributing to the opacity.

We have added WIMPs to models calibrated to the AGS05 abundances.  As we discuss below, WIMPs mainly would affect the innermost 10\% of the Sun's radius that is sampled least well by the observed solar p-modes \cite{Aerts_2010}.  
For the models presented here, the effects become noticeable only for $m_{\chi}\lesssim$\,20\gev \,when the WIMPs are less tightly bound in their orbits around the solar centre and can transfer energy to larger distances from the core.  
With decreasing WIMP mass, WIMP energy transport cools and thermalizes the core to lower temperatures out to larger radii.  Because the calibrated models need to remain in hydrostatic equilibrium and to generate the same luminosity, $\rho_c$ and the central hydrogen abundance increase to compensate for the cooler temperatures.  
The lower temperatures reduce the predicted neutrino flux, particularly the $^8$B $\nu$ flux, which has a steep temperature dependence of $T^{25}$ near the solar center \cite{Bahcall_2001}.  While the current solar neutrino experiments can accommodate the $^8$B neutrino flux predicted by standard solar models using either the GN93 or AGS05 abundances, more work is required to determine whether they can accommodate a flux as low as predicted for the discussed solar models including WIMPs.

With increasing density concentration in the {\rm solar} core, the envelope becomes less condensed, and would normally have a larger radius. Therefore, the mixing length $\alpha$ is slightly increased to calibrate the model to the observed solar radius, resulting in increased convective efficiency and the onset of convection occurring at a slightly smaller radius.  
This change in $\alpha$ is nevertheless much too small to deepen the convection zone base radius $R_{\rm CZB}$, while retaining the AGS05 abundances, to the value of 0.713 $\pm$ 0.001 $R_{\odot}$ determined from helioseismic data.

%%%%%%%%%%%%%%%%%%%%%%%%%%%%%%%%%%%%%%%%%%%%%%%
\begin{figure}[t]
\includegraphics[width=9cm]{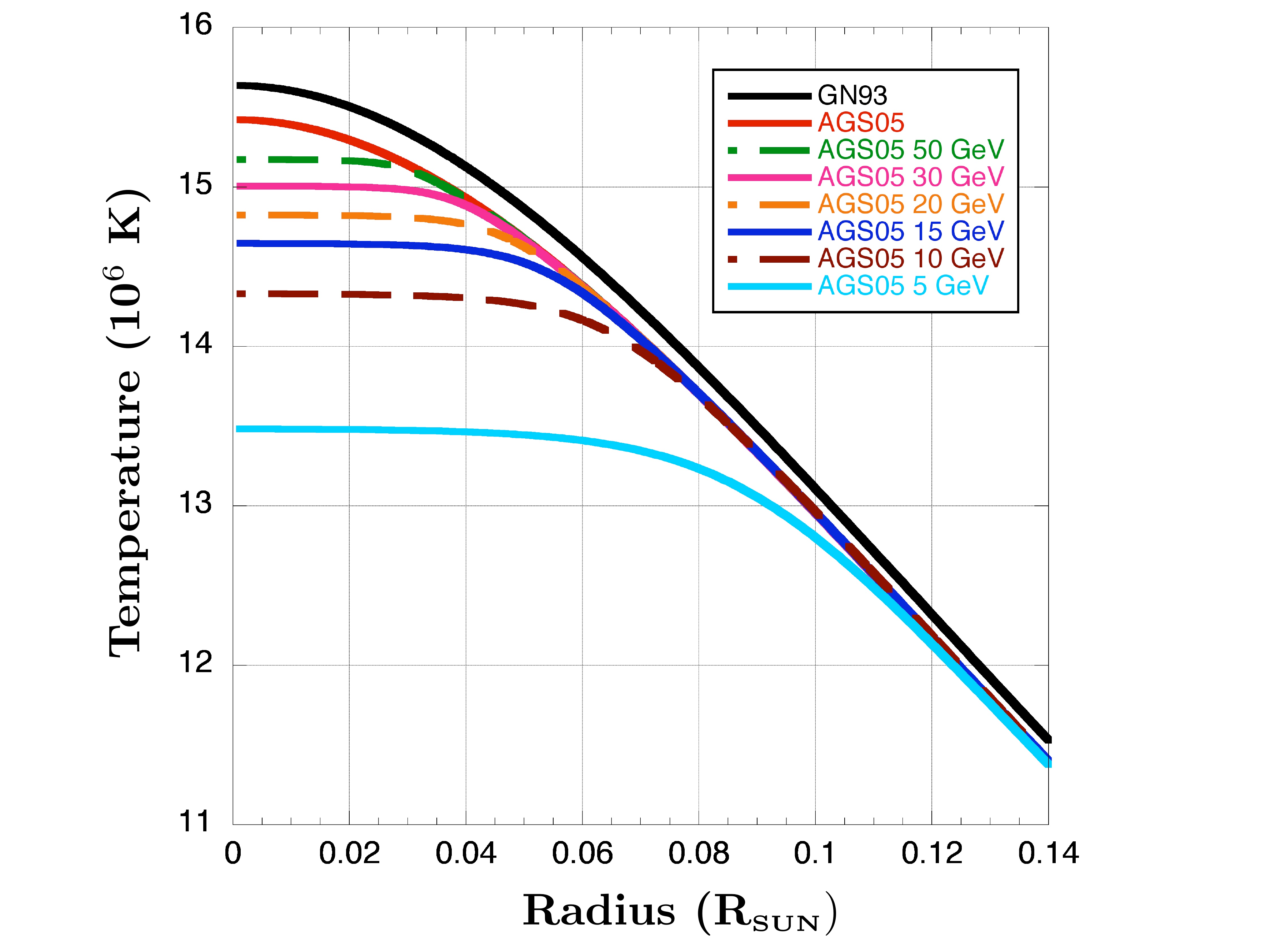}
\caption
{Central temperature versus fractional solar radius for solar models with WIMPS with masses 5-50 GeV compared to the AGS05 and GN93 model temperature profile. All WIMP models were run with $\langle\sigma_{\rm ann.}\upsilon\rangle=10^{-40}\,{\rm cm}^2$   (annihilations suppressed) and a spin-dependent cross section $\sigma^{\rm  sd}_{\chi p}=7\times 10^{-35}\,{\rm cm}^2$, using the AGS05 element abundances.}
 \label{fig:fig1}
\end{figure}
%%%%%%%%%%%%%%%%%%%%%%%%%%%%%%%%%%%%%%%%%%%%%%%

In Fig.\,\ref{fig:fig1}, we plot the solar temperature as a function of fractional solar radius, for solar models with WIMPs with masses $m_{\chi}=5-50\,\gev$, compared with the temperature profiles corresponding to the AGS05 and GN93 models.
As discussed above, the average orbital radius and interaction region of the WIMPs increases with decreasing mass. As seen in Table\,\ref{tab:maxevo}, the WIMPs reduce the effective opacity in the core to only approximately 0.02\,cm$^2$\,g$^{-1}$, compared to 1.2\,cm$^2$\,g$^{-1}$ for the standard solar models.  The transport of energy by the WIMPs is so efficient that the core essentially becomes isothermal out to the edge of the interaction region, where the temperature gradient approaches that of the standard model when WIMPs are omitted. For the most extreme 5\gev\,\,WIMP mass model, the solar temperature is significantly reduced out to a radius of approximately 0.1\,$R_{\odot}$.

%***************************************************************************************************

\subsection{Effect on sound speed}

%%%%%%%%%%%%%%%%%%%%%%%%%%%%%%%%%%%%%%%%%%%%%%%
\begin{figure}[t]
 \includegraphics[width=9cm]{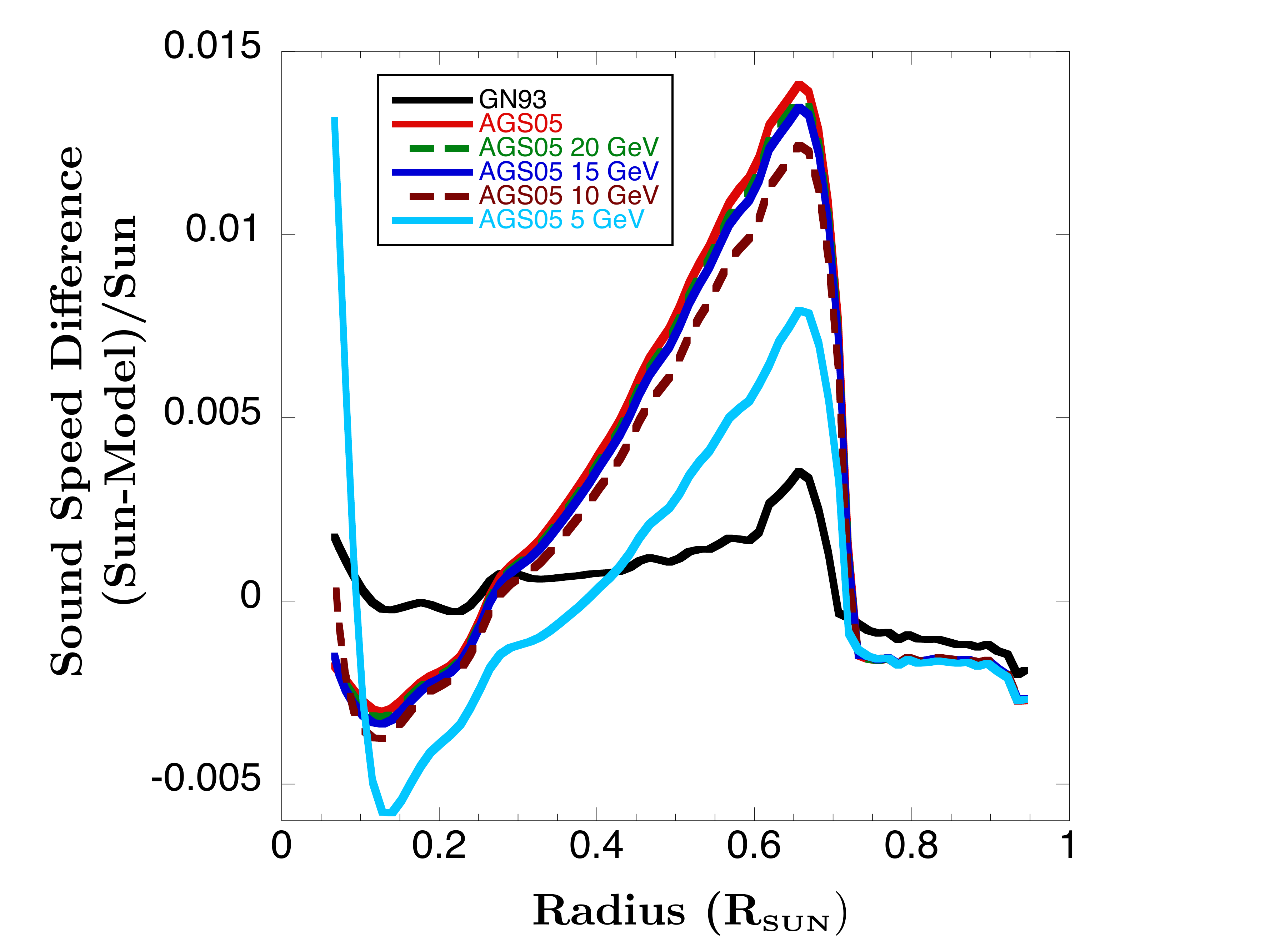}
 \caption{Relative difference between the inferred and calculated sound speeds for solar models constructed with the GN93 and AGS05 abundances.  Also compared are models with energy transport by WIMPs, based on the AGS05 abundances, for masses $m_x$ = 5,\,10,\,15 and 20\gev, $\langle\sigma_{\rm ann.}\upsilon\rangle\,=\,10^{-40}\,{\rm cm}^3\,{\rm s}^{-1}$ and a spin-dependent interaction cross section of $\sigma_x\,=\,7\times10^{-35}\,{\rm cm}^2$. 
The inferred sound speed is from \citep{basu00}. This combination of parameters enhances the effect WIMPs have on the sound-speed profile.}
\label{fig:fig2}
 \end{figure}
%%%%%%%%%%%%%%%%%%%%%%%%%%%%%%%%%%%%%%%%%%%%%%%

\noindent We now address the question as to whether the structural changes in the solar core discussed in the previous section lead to a detectable helioseismic signature.  Fig.\,\ref{fig:fig2} shows the difference between the sound speed inferred from solar p-modes, compared to that generated by standard models as well as those models including WIMPs with $5\le m_{\chi}\le20$\gev.  

Because only a few solar p-mode eigenfunctions of the lowest degree $0\le\ell\le2$ have significant amplitude near the solar core \cite{Aerts_2010}, the sound-speed inversions using p-modes are not sensitive to solar central conditions with high accuracy for radii within 0.06\,$R_{\odot}$ \cite{basu00}. In Fig.\,\ref{fig:fig2}, we omit plotting the results corresponding to $m_{\chi}=30$ and 50\gev, as they nearly coincide with those from the standard model with AGS05 abundances.  These sound-speed difference curves nearly coincide because such WIMPs only affect the model structure and sound speed profile for radii within 0.04\,$R_{\odot}$. 

From Fig.\,\ref{fig:fig2}, we observed that both 10\gev and 20\gev WIMPs have only a small effect on the sound-speed profile outside the central core at radii $0.06\lesssim r \lesssim0.2\,R_{\odot}$.  The models with WIMPs predict a lower central helium abundance (see Table\,\ref{tab:maxevo}) and a corresponding lower central mean molecular weight, $\mu$, while at the same time the temperature profile becomes identical to that of the standard model at radius $r\gtrsim$ 0.08\,$R_{\odot}$ for $m_{\chi}=10\gev$, and $\gtrsim$ 0.06\,$R_{\odot}$ for $m_{\chi}=20\gev$.   Therefore, the sound speed, which is proportional to $\sqrt{(T/\mu)}$, is increased in this region for the WIMP models.  
Unfortunately, the small increase in sound speed in this region, that could be diagnosed by sound speed inversions, is in the wrong direction to reduce the discrepancy with the seismic inversions  observed for the standard model AGS05 abundances.

%%%%%%%%%%%%%%%%%%%%%%%%%%%%%%%%%%%%%%%%%%%%%%%
 \begin{figure}[t]
 \includegraphics[width=10cm]{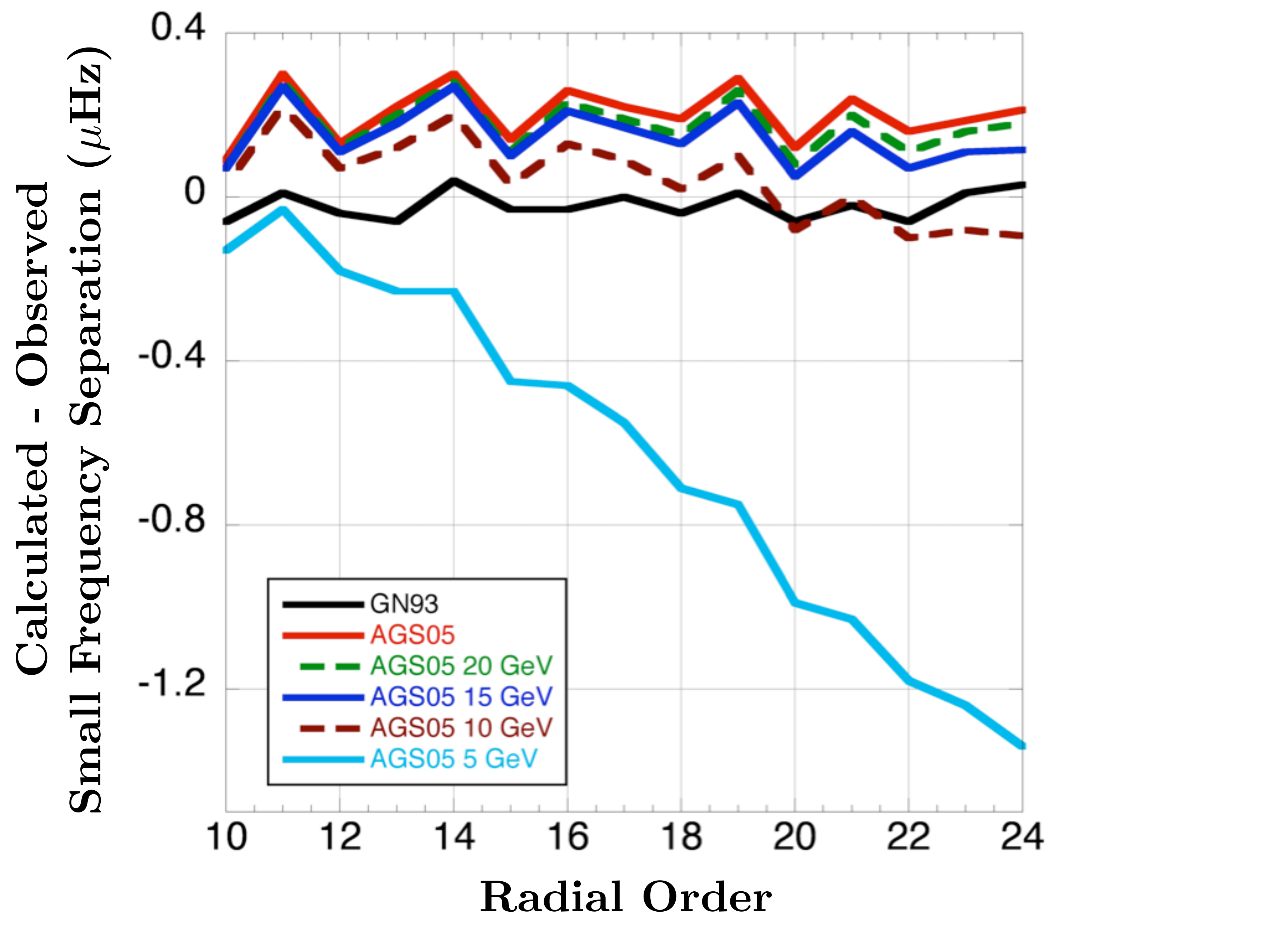}
 \caption{
 Difference between the calculated and observed small frequency separations for
p-modes of degree $\ell$ = 0 and $\ell$ = 2 against radial order for $\ell$ = 0 modes.
The observations are from \cite{chaplin07}, corrected for solar-cycle variations using the method described in \citep{basu07}.}
\label{fig:fig3}
\end{figure}
%%%%%%%%%%%%%%%%%%%%%%%%%%%%%%%%%%%%%%%%%%%%%%%

We also observe from Fig.\,\ref{fig:fig2}, that we can use the discrepancy in predictions for core sound-speed and p-mode frequencies of the solar model including 5\gev mass WIMPs to rule out this model.  
The large core temperature decrease produced by the WIMPs extends to 10\% of the solar radius, far enough out for p-mode sensitivity.  
The lower temperature results in a slower sound speed compared to that predicted by the standard AGS05 model for $r\lesssim0.1\,R_{\odot}$.  
When the temperature profile joins that of the AGS05 model at 0.1\,$R_{\odot}$, the much-reduced mean molecular weight results in an increase in sound speed compared to the standard AGS05 model at larger radii out to $0.4\,R_{\odot}$.   
Between 0.4\,$R_{\odot}$ and $R_{\rm CZB}=0.722\,R_{\odot}$, the 5\gev\,\,WIMP mass model actually mitigates the sound speed discrepancy for the AGS05 abundances, because the lower Y abundance of the calibrated model increases the sound speed, and the slightly larger mixing length/pressure-scale-height ratio improves the efficiency of convection and slightly deepens the convection zone. 
However, for this model and all of the AGS05 abundance models presented here with/without WIMPs, the Y abundance needed to calibrate the model to the solar luminosity is too low compared to that derived from the signature of helium ionization in the convection zone \cite{BA04a}, and the convection zones of these models are also still too shallow compared to the seismically-derived value.

%***************************************************************************************************
\subsection{Effect on small frequency separations}

\noindent The differences between the very precisely measured $\ell=0$ and  $\ell=2$ p-mode frequencies offset by one radial order, known as the {\it small separations}, are somewhat more sensitive to the solar central conditions.   
 
Fig.\,\ref{fig:fig3} shows the difference between the calculated and observed small separations.  Notice that the calculated small separations for the standard model with GN93 abundances agree well with observations, whereas the separations for the model with the AGS05 abundances are about 0.2 $\mu$Hz too high.  
There is very little effect on these separations for the AGS05 abundance models including WIMPs with mass $m_{\chi}>10\gev$.  The effect becomes more pronounced for the 10\gev\,\,WIMP mass model, where the WIMPs introduce a decreasing slope with increasing radial order.  
This slope becomes much more extreme for the 5\gev\,\,model, and consequently we find that helioseismic observations would rule out such a model.

%***************************************************************************************************

\subsection{Effect on gravity modes}
\noindent While the sound speed inferred from solar pressure modes and other solar structure properties outside the solar core are not very sensitive to the effects of WIMPs, the longer-period solar gravity modes, which have largest amplitudes in the solar core but very small amplitude near the solar surface \cite{Aerts_2010}, show more sensitivity.  To date, only one or at most a few g-modes have been possibly detected in long-term observations by the SOHO spacecraft \cite{Garcia:2008mi, Appourchaux:2009fe}. 
Table\,\ref{tab:gmodemax} lists the calculated g-mode frequencies for the standard (i.e., without WIMPs) solar models with GN93 and AGS05 abundances, and the calculated differences between the frequencies of the AGS05 model and the frequencies of the models with WIMP masses of 20, 10, and 5 GeV.  The g-modes with the lowest-degree $\ell$ have the largest predicted amplitudes at the solar surface, and therefore are the most likely to be detected.  
As can be observed from Fig.\,\ref{fig:fig3}, the changes in frequency when WIMPs of these masses are included are several to tens of $\mu$Hz.   
If g-mode frequencies are isolated and the modes are identified unambiguously, the frequencies should be measurable to less than a $\mu$Hz, and so will easily allow one to distinguish between solar models with/without WIMPs, or between WIMP models with different values of $m_{\chi}$.

%%%%%%%%%%%%%%%%%%%%%%%%%%%%%%%%%%%%%%%%%%%%%%%
\begin{table*}[t]
\begin{minipage}{\textwidth}
\caption[Caption]%
{Comparison of GN93, AGS05 and WIMP Model {\it g}-Mode Predictions\footnote[3]{
Typical low-order $g$-mode predictions for the GN93 and AGS05 models, and predicted difference from the AGS05 model prediction for solar models with WIMPs of $m_x$ = 20, 10, and 5 GeV.  All WIMP models were run with $\left<\sigma_{\rm ann}v\right>$ = 1 x 10$^{-40}$ cm$^3$ s$^{-1}$ and $\sigma_x$ = 7 x 10$^{-35}$ cm$^2$ in order to enhance the effects on the solar model.  The units of $\nu$ and the quantity  $\Delta {\nu}^{m_x}\,=\,\nu_{n,\ell}^{m_x}\,-\,\nu_{n,\ell}^{\rm AGS05}\;(n=-1,-5;\;\ell=1,6)$ are $\mu$Hz.
}}
\begin{center}
\begin{tabular}{lr|ccccc|lr|ccccc}
\noalign{\smallskip} \hline\hline
$\ell$ & $n$ & $\nu^{\rm GN93}$ & $\nu^{\rm AGS05}$ & $\Delta {\nu}^{20}$ & $\Delta {\nu}^{10}$  & $\Delta {\nu}^{5}$ &
$\ell$ & $n$ &  $\nu^{\rm GN93}$ & $\nu^{\rm AGS05}$ & $\Delta {\nu}^{20}$ & $\Delta {\nu}^{10}$ & $\Delta {\nu}^{5}$\\
\hline
1&-1& 260.67 & 257.25 & 4.64 & 13.67& 21.00 & 4&-1& 365.88 & 361.37& 2.75 & 12.84 & 21.60\\
1&-2& 189.63 & 186.21& 5.66& 11.05& 23.71 & 4& -2 & 325.13 &  320.29& 5.52 & 16.59 & 35.79\\
1&-3& 151.81 & 149.19& 4.17&  7.56& 15.20 & 4& -3 & 288.72 & 284.68& 6.10 & 13.89 & 32.79\\
1&-4& 126.48 & 124.11& 3.30&  6.91& 15.20 & 4 & -4 & 262.68 & 260.96& 4.18 & 9.43 & 24.38\\
1&-5& 108.06 & 105.94& 2.83&  5.85& 13.18 & 4& -5& 247.87 & 245.79& 4.18 & 7.86 & 14.39\\
\hline
2&-1& 295.70 & 293.25& 2.88&  9.34 & 20.93 & 5&-1& 383.19 & 378.67& 2.18 & 11.88 & 34.41\\
2&-2& 258.89 & 256.53& 4.61& 10.30& 21.35 & 5&-2& 347.81 & 342.71& 4.80 & 17.07 & 36.18\\
2&-3& 225.23 & 222.42& 5.21&  9.68& 21.46 & 5&-3& 313.45 & 308.79& 6.04 & 15.43 & 36.31\\
2&-4& 194.91 & 191.65& 5.51& 10.03& 22.08 &  5&-4& 285.57 & 282.79& 5.05 & 12.56 & 30.79\\
2&-5& 170.16 & 167.01& 4.41&  9.13& 20.32 & 5&-5& 270.50 & 270.06& 2.05 & 4.70 & 17.52\\
\hline
3&-1& 337.72 & 333.43& 3.41& 13.20& 30.10 & 6&-1& 394.36 & 389.83& 1.79 & 11.15 & 37.96\\
3&-2& 293.62 & 289.50& 5.84& 14.67& 31.98 & 6&-2& 364.26 & 359.03& 4.11 & 16.84 & 34.84\\
3&-3& 257.91 & 255.28& 5.21& 10.80& 26.07& 6&-3& 332.74 & 327.84& 5.50 & 16.02 & 38.09\\
3&-4& 233.77 & 232.06& 4.43&  8.76& 19.76 & 6&-4& 305.80 & 301.41& 6.29 & 15.06 & 34.86\\
3&-5& 213.36 & 210.60& 4.86&  9.69& 20.19 & 6&-5& 287.87 & 287.90& 1.16 & 5.42 & 22.82\\
\hline
\end{tabular}
\end{center}
\normalsize
\label{tab:gmodemax}
\end{minipage}
\end{table*}
%%%%%%%%%%%%%%%%%%%%%%%%%%%%%%%%%%%%%%%%%%%%%%%

%***************************************************************************************************

\section{Discussion and Conclusions}

Our main conclusion is that both direct detection and accelerator probes may be complemented by using the Sun as a probe of DM.  Models of DM that have large spin-dependent interactions and an intrinsic asymmetry that prevents post freeze-out annihilations can significantly lower the central temperature of the Sun as well as the resulting $^8$B neutrino flux.  For WIMP masses $m_{\chi}>10\gev$, the presence of WIMPs does not significantly affect currently available helioseismic constraints.  However, for WIMP masses of 10\,GeV or lighter, constraints on sound speed and small frequency separations between $\ell=0$ and $\ell=2$ p-modes can begin to constrain and rule out the presence of WIMPs with the cross sections utilised here.

Our study is motivated in a large part by the recently revised solar abundances \cite{AGS05, AGSS09} which result in solar models that cannot reproduce the currently observed helioseismic data, with numerous attempts to restore agreement being met with only partial success (see e.g., \cite{BA08, GM10}).  This means that additional physics must be incorporated into solar modelling, and dark matter is among the options that merit detailed consideration. 

Since the original submission of our paper in May 2010, an additional paper appeared on solar models including WIMPs and the implications for reconciling the new solar abundances with helioseismology \cite{Taoso10}.  In agreement with \cite{Taoso10}, our explorations to date do not show any realistic path in which the inclusion of WIMPs will mitigate this problem.  Even for the large interaction cross section and small annihilation cross section considered here, the inclusion of WIMPs of mass $m_{\chi}>10\gev$ has little effect on presently observable helioseismic signatures.  The inclusion of WIMPs with masses of 10\,GeV or lighter worsens the agreement with the helioseismically-inferred sound speed at radii $0.1\le r\le0.2\,R_{\odot}$, only slightly deepens the predicted convection-zone depth, and introduces a trend with radial order in the low-degree p-mode small separations that is not observed in the data. Our primary new result is that WIMP masses of $\sim 5$ GeV may be excluded for spin-dependent interactions in a specified cross-section range, thereby complementing direct detection experiments in a region that they access only with great difficulty provided the WIMPs annihilation cross section is suppressed.

While here we do not discuss whether these WIMP models could accommodate measurements of the $^8$B neutrino flux from current solar neutrino experiments such as Super-Kamiokande III \cite{Yang:2009hp}, SNO \cite{Collaboration:2009qz} or Borexino \cite{Collaboration:2008mr}, with precisions of $\sim$10\% and theoretical expectations of up to $\sim$20\% depending on the solar composition \cite{Bottino, Frandsen}, a more detailed study of the low-mass region of WIMP parameter space and its consistency with current experimental data is deferred to a later paper. 
There we will address the question of whether future helioseismic observations, most notably using g-modes, and solar neutrinos, may be able to constrain the allowable DM parameter space in a mass range that is of current interest for direct detection.

Finally, we note that for solar mass stars near the centre of the Galaxy, where the WIMP density is enhanced by up to some 6 orders of magnitude relative to that in the solar neighbourhood, the effect  of the redistribution of energy in the stellar core may generate a significant reduction of the main-sequence lifetimes.  We leave an investigation of this scenario to our future work.

%***************************************************************************************************

\section{acknowledgements}
\noindent DTC is supported by the Science and Technology Facilities Council. SMW thanks the Oxford physics department for hospitality and the Higher Education Funding Council for England and the Science and Technology Facilities Council for financial support under the SEPNet Initiative.

%***************************************************************************************************


\begin{thebibliography}{10}
\vspace{3mm}
\bibitem{Bottino}
  A.~Bottino, G.~Fiorentini, N.~Fornengo, B.~Ricci, S.~Scopel and F.~L.~Villante,
  %``Does solar physics provide constraints to weakly interacting massive
  %particles?,''
  Phys.\ Rev.\  D {\bf 66}, 053005 (2002)
  [arXiv:hep-ph/0206211].
  %%CITATION = PHRVA,D66,053005;%%

 %\cite{Lopes:2001ig}
\bibitem{lopesa}
  I.~Lopes and J.~Silk,
  %``Solar neutrinos: Probing the quasi-isothermal solar core produced by  SUSY
  %dark matter particles,''
  Phys.\ Rev.\ Lett.\  {\bf 88}, 151303 (2002)
  [arXiv:astro-ph/0112390].
  %%CITATION = PRLTA,88,151303;%%

%\cite{Lopes:2001ra}
\bibitem{lopesb}
  I.~P.~Lopes, J.~Silk and S.~H.~Hansen,
  %``Helioseismology as a new constraint on SUSY dark matter,''
  Mon.\ Not.\ Roy.\ Astron.\ Soc.\  {\bf 331} (2002) 361
  [arXiv:astro-ph/0111530].
  %%CITATION = MNRAA,331,361;%%


%\cite{Lopes:2002gp}
\bibitem{lopesc}
  I.~P.~Lopes, G.~Bertone and J.~Silk,
  %``Solar seismic model as a new constraint on supersymmetric dark matter,''
  Mon.\ Not.\ Roy.\ Astron.\ Soc.\  {\bf 337} (2002) 1179
  [arXiv:astro-ph/0205066].
  %%CITATION = MNRAA,337,1179;%%


 %\cite{Angloher:2002in}
\bibitem{cresst}
  G.~Angloher {\it et al.},
  %``Limits on WIMP dark matter using sapphire cryogenic detectors,''
  Astropart.\ Phys.\  {\bf 18} (2002) 43.
  %%CITATION = APHYE,18,43;%% 

%\cite{Angle:2007uj}
\bibitem{xenon10}
  J.~Angle {\it et al.}  [XENON Collaboration],
  %``First Results from the XENON10 Dark Matter Experiment at the Gran Sasso
  %National Laboratory,''
  Phys.\ Rev.\ Lett.\  {\bf 100} (2008) 021303
  [arXiv:0706.0039 [astro-ph]].
  %%CITATION = PRLTA,100,021303;%%
  
%\cite{Aprile:2010um}
\bibitem{xenon100}
  E.~Aprile {\it et al.}  [XENON100 Collaboration],
  %``First Dark Matter Results from the XENON100 Experiment,''
  Phys.\ Rev.\ Lett.\  {\bf 105}, 131302 (2010)
  [arXiv:1005.0380 [astro-ph.CO]].
  %%CITATION = PRLTA,105,131302;%%
  
\bibitem{Behnke}
E.~Behnke {\it et al.}  [COUPP Collaboration],
[arXiv1008.3518 [astro-ph]].

%\cite{Archambault:2009sm}
\bibitem{picasso}
  S.~Archambault {\it et al.},
  %``Dark Matter Spin-Dependent Limits for WIMP Interactions on 19-F by
  %PICASSO,''
  Phys.\ Lett.\  B {\bf 682} (2009) 185
  [arXiv:0907.0307 [hep-ex]].
  %%CITATION = PHLTA,B682,185;%%
 
%\cite{Behnke:2008zza}
\bibitem{coupp}
  E.~Behnke {\it et al.}  [COUPP Collaboration],
  %``Improved Spin-Dependent WIMP Limits from a Bubble Chamber,''
  Science {\bf 319}, 933 (2008)
  [arXiv:0804.2886 [astro-ph]].
  %%CITATION = SCIEA,319,933;%%

%\cite{Green:2010ri}
\bibitem{green}
  A.~M.~Green,
  %``Extracting information about WIMP properties from direct detection
  %experiments: astrophysical uncertainties,''
  arXiv:1004.2383 [astro-ph.CO].
  %%CITATION = ARXIV:1004.2383;%%

%\cite{Bernabei:2010mq}
\bibitem{dama}
  R.~Bernabei {\it et al.},
  %``New results from DAMA/LIBRA,''
  Eur.\ Phys.\ J.\  C {\bf 67}, 39 (2010)
  [arXiv:1002.1028 [astro-ph.GA]].
  %%CITATION = EPHJA,C67,39;%%

%\cite{Ahmed:2009zw}
\bibitem{cdms}
  Z.~Ahmed {\it et al.}  [The CDMS-II Collaboration],
  %``Dark Matter Search Results From The Cdms Ii Experiment,''
  Science {\bf 327}, 1619 (2010)
  [arXiv:0912.3592 [astro-ph.CO]].
  %%CITATION = SCIEA,327,1619;%%
  
%\cite{Aalseth:2010vx}
\bibitem{cogent}
  C.~E.~Aalseth {\it et al.}  [CoGeNT collaboration],
  %``Results from a Search for Light-Mass Dark Matter with a P-type Point
  %Contact Germanium Detector,''
  arXiv:1002.4703 [astro-ph.CO].
  %%CITATION = ARXIV:1002.4703;%%

\bibitem{lowmasssector}
%\cite{Fitzpatrick:2010em}
%\bibitem{Fitzpatrick:2010em}
  A.~L.~Fitzpatrick, D.~Hooper and K.~M.~Zurek,
  %``Implications of CoGeNT and DAMA for Light WIMP Dark Matter,''
  Phys.\ Rev.\  D {\bf 81}, 115005 (2010);
  %[arXiv:1003.0014 [hep-ph]];
  %%CITATION = PHRVA,D81,115005;%%
%\cite{Kuflik:2010ah}
%\bibitem{Kuflik:2010ah}
  E.~Kuflik, A.~Pierce and K.~M.~Zurek,
  %``Light Neutralinos with Large Scattering Cross Sections in the Minimal
  %Supersymmetric Standard Model,''
  Phys.\ Rev.\  D {\bf 81}, 111701 (2010);
  %[arXiv:1003.0682 [hep-ph]];
  %%CITATION = PHRVA,D81,111701;%%
%\cite{Andreas:2010dz}
%\bibitem{Andreas:2010dz}
  S.~Andreas, C.~Arina, T.~Hambye, F.~S.~Ling and M.~H.~G.~Tytgat,
  %``A light scalar WIMP through the Higgs portal and CoGeNT,''
  Phys.\ Rev.\  D {\bf 82}, 043522 (2010);
  %[arXiv:1003.2595 [hep-ph]];
  %%CITATION = PHRVA,D82,043522;%%
%\cite{Chang:2010yk}
%\bibitem{Chang:2010yk}
  S.~Chang, J.~Liu, A.~Pierce, N.~Weiner and I.~Yavin,
  %``CoGeNT Interpretations,''
  JCAP {\bf 1008}, 018 (2010);
  %[arXiv:1004.0697 [hep-ph]];
  %%CITATION = JCAPA,1008,018;%%
%\cite{Graham:2010ca}
%\bibitem{Graham:2010ca}
  P.~W.~Graham, R.~Harnik, S.~Rajendran and P.~Saraswat,
  %``Exothermic Dark Matter,''
  arXiv:1004.0937 [hep-ph];
  %%CITATION = ARXIV:1004.0937;%%
%\cite{Foot:2010rj}
%\bibitem{Foot:2010rj}
  R.~Foot,
  %``A CoGeNT confirmation of the DAMA signal,''
  Phys.\ Lett.\  B {\bf 692}, 65 (2010);
  %[arXiv:1004.1424 [hep-ph]];
  %%CITATION = PHLTA,B692,65;%%
%\cite{An:2010kc}
%\bibitem{An:2010kc}
  H.~An, S.~L.~Chen, R.~N.~Mohapatra, S.~Nussinov and Y.~Zhang,
  %``Energy Dependence of Direct Detection Cross Section for Asymmetric Mirror
  %Dark Matter,''
  Phys.\ Rev.\  D {\bf 82}, 023533 (2010);
  %[arXiv:1004.3296 [hep-ph]];
  %%CITATION = PHRVA,D82,023533;%%
D.~T.~Cumberbatch, D.~E.~Lopez-Fogliani, R.~Ruiz de Austri, 
L.~Roszkowski, Y.~S.~Tsai and T.~Varley
{\it in preparation}.



%\cite{Kopp:2009qt}
\bibitem{Kopp}
  J.~Kopp, T.~Schwetz and J.~Zupan,
  %``Global interpretation of direct Dark Matter searches after CDMS-II
  %results,''
  JCAP {\bf 1002}, 014 (2010)
  [arXiv:0912.4264 [hep-ph]].
  %%CITATION = JCAPA,1002,014;%%
  
  
%\cite{Serenelli:2009yc}
\bibitem{serenelli}
  A.~Serenelli, S.~Basu, J.~W.~Ferguson and M.~Asplund,
  %``New Solar Composition: The Problem With Solar Models Revisited,''
  2009 ApJ 705 L123,
  arXiv:0909.2668 [astro-ph.SR].
  %%CITATION = ARXIV:0909.2668;%%
    
%\cite{Asplund:2004eu}
\bibitem[{Asplund {\it et al.} (2005)}]{AGS05}
  M.~Asplund, N.~Grevesse and J.~Sauval,
  %``The solar chemical composition,''
  Nucl.\ Phys.\  A {\bf 777} (2006) 1
  [ASP Conf.\ Ser.\  {\bf 336} (2005) 25]
  [arXiv:astro-ph/0410214].
  %%CITATION = ASPSF,336,25;%%

%\cite{Asplund:2009fu}
\bibitem[{Asplund et al.(2009)}]{AGSS09}
  M.~Asplund, N.~Grevesse, A.~J.~Sauval and P.~Scott,
  %``The chemical composition of the Sun,''
  Ann.\ Rev.\ Astron.\ Astrophys.\  {\bf 47} (2009) 481
  [arXiv:0909.0948 [astro-ph.SR]].
  %%CITATION = ARAAA,47,481;%%

   %\cite{Frandsen:2010yj}
\bibitem{Frandsen}
  M.~T.~Frandsen and S.~Sarkar, Phys. Rev. Lett. 105, 011301 (2010).
  %``Asymmetric dark matter and the Sun,''
  %arXiv:1003.4505 [hep-ph].
  %%CITATION = ARXIV:1003.4505;%%
  
 
 %\cite{Gondolo:2004sc}
\bibitem{Gondolo:2004sc}
  P.~Gondolo, J.~Edsjo, P.~Ullio, L.~Bergstrom, M.~Schelke and E.~A.~Baltz,
  %``DarkSUSY: Computing supersymmetric dark matter properties numerically,''
  JCAP {\bf 0407}, 008 (2004)
  [arXiv:astro-ph/0406204].
  %%CITATION = JCAPA,0407,008;%%

 
 %\cite{Bertone:2004pz}
\bibitem{bhs}
  G.~Bertone, D.~Hooper and J.~Silk,
  %``Particle dark matter: Evidence, candidates and constraints,''
  Phys.\ Rept.\  {\bf 405} (2005) 279
  [arXiv:hep-ph/0404175].
  %%CITATION = PRPLC,405,279;%%
  
  %\cite{Bass:2004xa}
\bibitem{Bass:2004xa}
  S.~D.~Bass,
  %``The spin structure of the proton,''
  Rev.\ Mod.\ Phys.\  {\bf 77}, 1257 (2005)
  [arXiv:hep-ph/0411005].
  %%CITATION = RMPHA,77,1257;%%
  
  
 \bibitem{asymm}
  D.~B.~Kaplan,
  %``A Single explanation for both the baryon and dark matter densities,''
  Phys.\ Rev.\ Lett.\  {\bf 68}, 741 (1992);
  %%CITATION = PRLTA,68,741;%%
 %
%\bibitem{Hooper:2004dc}
  D.~Hooper, J.~March-Russell and S.~M.~West,
  %``Asymmetric sneutrino dark matter and the Omega(b)/Omega(DM) puzzle,''
  Phys.\ Lett.\  B {\bf 605}, 228 (2005);
  %[arXiv:hep-ph/0410114].
  %%CITATION = PH
  %\cite{Kitano:2004sv}
%\bibitem{Kitano:2004sv}
  R.~Kitano and I.~Low,
  %``Dark matter from baryon asymmetry,''
  Phys.\ Rev.\  D {\bf 71}, 023510 (2005)
  [arXiv:hep-ph/0411133];
  %%CITATION = PHRVA,D71,023510;%%LTA,B605,228;%%
%
%\bibitem{Kaplan:2009ag}
  D.~E.~Kaplan, M.~A.~Luty and K.~M.~Zurek,
  %``Asymmetric Dark Matter,''
  Phys.\ Rev.\  D {\bf 79}, 115016 (2009);
  %[arXiv:0901.4117 [hep-ph]].
  %%CITATION = PHRVA,D79,115016;%%
%
%\bibitem{Kribs:2009fy}
  G.~D.~Kribs, T.~S.~Roy, J.~Terning and K.~M.~Zurek,
  %``Quirky Composite Dark Matter,''
  arXiv:0909.2034. [hep-ph].
  %%CITATION = ARXIV:0909.2034;%%

  
\bibitem{gr}
  G.~F.~Giudice and S.~Raby,
  %``Cosmions with spin-dependent interactions,''
  Phys.\ Lett.\  B {\bf 247} (1990) 423;
  %%CITATION = PHLTA,B247,423;%%
A.~N.~Cox, J.~A.~Guzik and S.~Raby, Astrophys.\ J.\  {\bf 353}, 698 (1990).
%%CITATION = LA-UR-89-1152;%%

\bibitem{solarneuprob}
%\bibitem{Faulkner:1985rm}
  J.~Faulkner and R.~L.~Gilliland,
  %``Weakly interacting, massive particles and the solar neutrino flux,''
  Astrophys.\ J.\  {\bf 299}, 994 (1985);
  %%CITATION = ASJOA,299,994;%%
%\bibitem{Spergel:1984re}
  D.~N.~Spergel and W.~H.~Press,
  %``Effect of hypothetical, weakly interacting, massive particles on energy
  %transport in the solar interior,''
  Astrophys.\ J.\  {\bf 294}, 663 (1985);
  %%CITATION = ASJOA,294,663;%%
%\bibitem{Press:1985ug}
  W.~H.~Press and D.~N.~Spergel,
  % ``Capture by the Sun of a galactic population of weakly
  % interacting, massive particles,''
  Astrophys.\ J.\  {\bf 296}, 679 (1985);
  %%CITATION = ASJOA,296,679;%%
%\bibitem{Gilliland:1986}
   R.~L.~Gilliland, J.~Faulkner, W.~H.~Press and D.~N.~Spergel,
   %``Solar models with energy transport by weakly interacting particles,''
   Astrophys.\ J.\  {\bf 306}, 703 (1986);
  %%CITATION = ASJOA,306,703;%%
  G.~G.~Ross and G.~C.~Segre,
  %``A Broken E(6) Solution To The Solar Neutrino Problem,''
  Phys.\ Lett.\  B {\bf 197}, 45 (1987).
  %%CITATION = PHLTA,B197,45;%%
  
%\cite{Gould:1987ju}
\bibitem{gould1987}
  A.~Gould,
  %``WIMP DISTRIBUTION IN AND EVAPORATION FROM THE SUN,''
  Astrophys.\ J.\  {\bf 321}, 560 (1987).
  %%CITATION = ASJOA,321,560;%%

%\cite{Griest:1986yu}
\bibitem{griest1987}
  K.~Griest and D.~Seckel,
  %``Cosmic Asymmetry, Neutrinos and the Sun,''
  Nucl.\ Phys.\  B {\bf 283} (1987) 681
  [Erratum-ibid.\  B {\bf 296} (1988) 1034].
  %%CITATION = NUPHA,B283,681;%%


\bibitem[Iben(1963)]{Iben_1963}
I.~Iben 1963, \apj, 138,452.

\bibitem[Iben(1965a)]{Iben_1965a}
I.~Iben, 1965, \apj, 141, 993.

\bibitem[Iben(1965b)]{Iben_1965b}
I.~Iben, 1965, \apj, 142, 1447.

\bibitem[Guzik \& Swenson(1997)]{GS97}
J.~A.~Guzik, J.A. \&  F.~J.~Swenson, 1997, \apj, 491, 967.

\bibitem[Burgers(1969)]{burgers69} 
J.M. Burgers, 1969, Flow Equations for Composite Gases, (New York: Academic).

\bibitem[{Cox {\it et al.}(1989)}]{CGK89}
A.N. Cox,  J.A. Guzik, \& R.B. Kidman, 1989, \apj, 342, 1187 (CGK89).

\bibitem[{Iben \& McDonald (1985}]{IM85}
I. Iben, Jr. \& MacDonald, J. 1985, \apj, 296, 540.

\bibitem[{Angulo {\it et al.}(1999)}]{Angulo_1999}
C. Angulo,  {\it et al.} 1999, Nucl. Phys A, 656, 3.
  
\bibitem[Formicola {\it et al.}(2004)]{Formicola_2004}
A. Formicola, {\it et al.} (2004), Phys. Lett. B, 591, 61.
  

%\cite{Iglesias:1996bh}
\bibitem{IR96}
  C.~A.~Iglesias and F.~J.~Rogers,
  %``Updated Opal Opacities,''
  Astrophys.\ J.\  {\bf 464} (1996) 943.
  %%CITATION = ASJOA,464,943;%%


\bibitem[{Ferguson {\it et al.}(2005)}]{Fer05}
J.W. Ferguson,  D.R. Alexander,  F. Allard, T. Barman, J.G. Bodnarik, P.H. Hauschildt, A. Heffner-Wong,  \&  A.  Tamanai, 2005, \apj, 623, 585.

\bibitem[{Allen(1973)}]{Allen_1973}
C.W. Allen, 1973, Astrophysical Quantities, 3rd Edition (London: Athlone Press), 169.
  
\bibitem[{Willson {\it et al.}(1986)}]{Willson_1986}
R.C. Willson, H.S. Hudson, C. Frohlich,  \& R.W. Brusa, 1986, Science, 234, 1114.
  
\bibitem[{Cohen \& Taylor(1986)}]{CT86}
E.R. Cohen, \& B.N. Taylor, 1986, Codata Bulletin, 63 (Boulder: NBS), 1.

\bibitem[Guenther {\it et al.}(1992)]{Guenther_1992}
D.B. Guenther, P. Demarque, Y.-C. Kim,  \& M.H. Pinsonneault,  1992, \apj, 387, 372.

\bibitem[{Grevesse \& Noels(1993)}]{GN93}
N. Grevesse, \& A. Noels, 1993, in Origin and Evolution of the Elements, Cambridge U. Press, 15 (GN93).
 
\bibitem[Pesnell(1990)]{Pesnell_1990}
W.D. Pesnell, 1990, \apj, 363, 227.
  

%\cite{Guzik:2005db}
\bibitem[{Guzik {\it et al.}(2005)}]{GWC05}
  J.~A.~Guzik, L.~S.~Watson and A.~N.~Cox,
  %``Can Enhanced Diffusion Improve Helioseismic Agreement for Solar Models with
  %Revised Abundances?,''
  Astrophys.\ J.\  {\bf 627} (2005) 1049
  [arXiv:astro-ph/0502364].
  %%CITATION = ASJOA,627,1049;%%


%\cite{Gould:1989hm}
\bibitem{GR90a}
  A.~Gould and G.~Raffelt,
  %``THERMAL CONDUCTION BY MASSIVE PARTICLES,''
  Astrophys.\ J.\  {\bf 352} (1990) 654.
  %%CITATION = ASJOA,352,654;%%

%\cite{Gould:1989ez}
\bibitem{GR90b}
  A.~Gould and G.~Raffelt,
  %``COSMION ENERGY TRANSFER IN STARS: THE KNUDSEN LIMIT,''
  Astrophys.\ J.\  {\bf 352} (1990) 669.
  %%CITATION = ASJOA,352,669;%%


\bibitem{WatsonDissertation}
L.S. Watson, Solar Models Including Revised Abundances and Dark Matter: Constraints from Helioseismology and Neutrino Observations, St. John's College, University of Oxford, March 2008.

 %\cite{Basu:2004zg}
\bibitem[Basu \& Antia(2004)]{BA04a}
  S.~Basu and H.~M.~Antia,
  %``Constraining solar abundances using helioseismology,''
  Astrophys.\ J.\  {\bf 606}, L85 (2004)
  [arXiv:astro-ph/0403485].
  %%CITATION = ASJOA,606,L85;%%


\bibitem[{Aerts {i\it et al.}(2010))}]{Aerts_2010}
 C. Aerts, J. Christensen-Dalsgaard,  \& D.W. Kurtz, 2010, Asteroseismology (Springer Astronomy and Astrophysics Library), p. 225-227.


%\cite{Bahcall:2001pe}
\bibitem{Bahcall_2001}
  J.~N.~Bahcall,
  %``How many sigma's is the solar neutrino effect?,''
  Phys.\ Rev.\  C {\bf 65}, 015802 (2002)
  [arXiv:hep-ph/0108147].
  %%CITATION = PHRVA,C65,015802;%%

%\cite{Basu:1999aa}
\bibitem[{Basu {\it et al.}(2000)}]{basu00}
  S.~Basu, M.~H.~Pinsonneault and J.~N.~Bahcall,
  %``How much do helioseismological inferences depend upon the assumed reference
  %model?,''
  Astrophys.\ J.\  {\bf 529} (2000) 1084
  [arXiv:astro-ph/9909247].
  %%CITATION = ASJOA,529,1084;%%
  
%\cite{Chaplin:2007uh}
\bibitem[{Chaplin {\it et al.}(2007)}]{chaplin07}
  W.~J.~Chaplin, A.~M.~Serenelli, S.~Basu, Y.~Elsworth, R.~New and G.~A.~Verner,
  %``Solar heavy element abundance: constraints from frequency separation
  %ratios of low-degree p modes,''
  Astrophys.\ J.\  {\bf 670} (2007) 872
  [arXiv:0705.3154 [astro-ph]].
  %%CITATION = ASJOA,670,872;%%
 
%\cite{Basu:2006vh}
\bibitem{basu07}
  S.~Basu, W.~J.~Chaplin, Y.~Elsworth, R.~New, A.~M.~Serenelli and G.~A.~Verner,
  %``Solar abundances and helioseismology: fine structure spacings and
  %separation ratios of low-degree p modes,''
  Astrophys.\ J.\  {\bf 655} (2007) 660
  [arXiv:astro-ph/0610052].
  %%CITATION = ASJOA,655,660;%%

%\cite{Garcia:2008mi}
\bibitem{Garcia:2008mi}
R.~A.~Garcia {\it et al.},
%``Update on g-mode research,''
Astron.\ Nachr.\  {\bf 329}, 476 (2008)
[arXiv:0802.4296 [astro-ph]];
%%CITATION = ASNAA,329,476;%%
%\cite{Garcia:2009uh}
%\bibitem{Garcia:2009uh}
R.~A.~Garcia,
%``The Sun as a Star: 13 years of SoHO,''
AIP Conf.\ Proc.\  {\bf 1170}, 560 (2009)
[arXiv:0907.4439 [astro-ph.SR]].
%%CITATION = APCPC,1170,560;%%

%\cite{Appourchaux:2009fe}
\bibitem{Appourchaux:2009fe}
  T.~Appourchaux {\it et al.},
  %``The quest for the solar g modes,''
  Astron. Astrophys.
Rev. (2010), 18, 197,
  arXiv:0910.0848 [astro-ph.SR].
  %%CITATION = ARXIV:0910.0848;%%
  
%\cite{Basu:2007fp}
\bibitem[Basu \& Antia(2008)]{BA08}
  S.~Basu and H.~M.~Antia,
  %``Helioseismology and Solar Abundances,''
  Phys.\ Rept.\  {\bf 457}, 217 (2008)
  [arXiv:0711.4590 [astro-ph]].
  %%CITATION = PRPLC,457,217;%%

%\cite{Guzik:2010ck}
\bibitem[Guzik \& Mussack(2010)]{GM10}
  J.~A.~Guzik and K.~Mussack,
  %``Exploring mass loss, low-Z accretion, and convective overshoot in solar
  %models to mitigate the solar abundance problem,''
  Astrophys.\ J.\  {\bf 713}, 1108 (2010)
  [arXiv:1001.0648 [astro-ph.SR]].
  %%CITATION = ASJOA,713,1108;%%

  %\cite{Taoso10}
\bibitem{Taoso10}
  M.~Taoso, F.~Iocco, G.~Meynet, G.~Bertone and P.~Eggenberger,
  %``Effect of low mass dark matter particles on the Sun,''
  arXiv:1005.5711 [astro-ph.CO].
  %%CITATION = ARXIV:1005.5711;%%
  
 
%\cite{Yang:2009hp}
\bibitem{Yang:2009hp}
  B.~Yang  [Super-Kamiokande Collaboration],
  %``Solar Neutrino Measurement at SK-III,''
  arXiv:0909.5469 [hep-ex].
  %%CITATION = ARXIV:0909.5469;%%

%\cite{Collaboration:2009qz}
\bibitem{Collaboration:2009qz}
  S.~Collaboration,
  %``Searches for High Frequency Variations in the $^8$B Solar Neutrino Flux at
  %the Sudbury Neutrino Observatory,''
  Astrophys.\ J.\  {\bf 710} (2010) 540
  [arXiv:0910.2433 [astro-ph.SR]].
  %%CITATION = ASJOA,710,540;%%

%\cite{Collaboration:2008mr}
\bibitem{Collaboration:2008mr}
 G.~Bellini {\it et al.}  [The Borexino Collaboration],
  %``Measurement of the solar 8B neutrino rate with a liquid scintillator target
  %and 3 MeV energy threshold in the Borexino detector,''
  Phys.\ Rev.\  D {\bf 82} (2010) 033006
  [arXiv:0808.2868 [astro-ph]].
  %%CITATION = PHRVA,D82,033006;%%

 \end{thebibliography}
 \end{document}